\documentclass{article}

\usepackage{arxiv}

\usepackage[utf8]{inputenc} % allow utf-8 input

\usepackage{amsfonts}       % blackboard math symbols
\usepackage{nicefrac}       % compact symbols for 1/2, etc.
\usepackage{microtype}      % microtypography
\usepackage{lipsum}

\usepackage{color}
\usepackage{listings}
\usepackage{booktabs}
\usepackage{array}
\usepackage{graphicx}
\usepackage{longtable}
\usepackage{subfigure}

\usepackage{cite}
\usepackage{url}
\usepackage{fancyhdr}

\usepackage{soul}

\usepackage{float}
\usepackage{listings}
\usepackage{mdwmath}
\usepackage{mdwtab}
\usepackage{multirow}
\usepackage{multicol}
\usepackage{rotating}
\usepackage{setspace}
\usepackage[utf8]{inputenc}
\usepackage{lineno}
\usepackage{listings}

\usepackage{mdwmath}
\usepackage{mdwtab}
\usepackage{multirow}
\usepackage{multicol}
\usepackage{array}
\usepackage{booktabs}

\usepackage{enumitem}
\usepackage{xspace}
\usepackage[export]{adjustbox}
\usepackage{graphicx}
\usepackage{color,soul}
\usepackage{rotating}
\usepackage{setspace}
\usepackage{amsmath} 
\usepackage{amssymb}
\usepackage{float}
\usepackage{xcolor}

\usepackage{hyperref}
\usepackage[numbers]{natbib}
\usepackage{url}

\title{A Pattern Language for Blockchain Governance}
\author{Yue Liu\textsuperscript{1,2}, Qinghua Lu\textsuperscript{1,2}, Guangsheng Yu\textsuperscript{1}, Hye-Young Paik\textsuperscript{2}, Harsha Perera\textsuperscript{1}, Liming Zhu\textsuperscript{1,2}\\
\textsuperscript{1}Data61, CSIRO, Australia\\
\textsuperscript{2}University of New South Wales, Australia\\
yue.liu@data61.csiro.au, qinghua.lu@data61.csiro.au,
saber.yu@data61.csiro.au\\
h.paik@unsw.edu.au,
harsha.perera@data61.csiro.au,
liming.zhu@data61.csiro.au}
\begin{document}

\maketitle

\begin{abstract}
Blockchain technology has been used to build next-generation applications taking advantage of its decentralised nature. Nevertheless, there are some serious concerns about the trustworthiness of blockchain due to the vulnerabilities in on-chain algorithmic mechanisms, and tedious disputes and debates in off-chain communities. Accordingly, blockchain governance has received great attention for improving the trustworthiness of all decisions that direct a blockchain platform. However, there is a lack of systematic knowledge to guide practitioners to perform blockchain governance. We have performed a systematic literature review to understand the state-of-the-art of blockchain governance. We identify the lifecycle stages of a blockchain platform, and present 14 architectural patterns for blockchain governance in this study. This pattern language can provide guidance for the effective use of patterns for blockchain governance in practice, and support the architecture design of governance-driven blockchain systems.

\end{abstract}

Blockchain, Governance, Architectural pattern, Incentive, Decision rights, Accountability

\section{Introduction}
Blockchain is an innovative distributed ledger technology that eliminates trusted third-parties in business processes. Blockchain enables agreements on transactional data sharing across a large network of untrusted participants~\cite{scheuermann2015iacr}. In recent years, a considerable number of projects and studies have been conducted, to explore how to leverage blockchain as a component in existing software architectures and workflows~\cite{2019-Bratanova-ACS}.

Blockchain has brought its unique characteristics (e.g. transparency, immutability, on-chain autonomy) into diverse usage scenarios. Nevertheless, there are significantly increased concerns on governance issues after two severe negative crises in two world-renowned blockchain platforms: Ethereum and Bitcoin. The Ethereum "DAO" attack in 2016 implied that flaws in smart contract code may affect the operation of on-chain algorithmic mechanisms. This was resolved by conducting a hard fork to reverse the impacted transactions and recover the stolen tokens (over 60 million dollars)~\cite{DAOattack}. In Bitcoin, the debate on block size resulted in the split of the whole ecosystem~\cite{BitcoinSize}. These events exposed the need for effective human oversight and coordination for both on-chain and off-chain businesses, and highlighted the importance of trustworthy governance processes for blockchain.

Blockchain governance refers to the structures and processes put in place to ensure that the development and use of blockchain are compliant with legal regulations and ethical responsibilities~\cite{liu2021systematic}. With the absence of a clear source of authority, existing governance frameworks (e.g., IT governance~\cite{weill2004governance, cobit2012business}, data governance~\cite{ballard2014ibm, ISO38505}) are not directly applicable to blockchain technology. Accordingly, this topic has received continuous attention from both academia and industry in recent years. For instance, there are researches focusing on customised governance methods in permissioned blockchains~\cite{selected1, selected14}, regulations of blockchain-based decentralised finance applications~\cite{selected3, selected16}, and high-level frameworks for blockchain governance~\cite{selected11, selected14, hofman2021blockchain}. In a real-world implementation, Dash introduces the concept of "masternodes", who can participate in significant decision-making processes \cite{dash_whitepaper}. Tezos enables smooth on-chain protocol replacement without a hard fork, to minimise the impacts on business processes~\cite{tezos_whitepaper}.

Nevertheless, there is a steep learning curve for practitioners to design and choose governance methods appropriate for their systems. Having systematic and holistic guidance can assist system architects and developers to determine who is involved in a governance method and when to perform the method. In this regard, this paper presents a pattern language for realising governance for blockchain. We analysed the lifecycle of a blockchain platform, and identified 18 architectural patterns for blockchain governance. Specifically, we present 14 patterns using the pattern form~\cite{patternLanguage}. The intended audience of this pattern language is software architects and developers, and researchers who are interested in the governance of blockchain. The pattern language can advance the design of governance-driven blockchain systems. To the best of our knowledge, this is the first paper presenting patterns for blockchain governance.

The remainder of this paper is organised as follows. Section~\ref{background} introduces background knowledge and related work. The overview of our pattern language is illustrated in Section~\ref{methodology}. Section~\ref{patterns} presents each pattern in details. Section~\ref{conclusion} concludes the paper.

\section{Background and Related Work}
\label{background}

\subsection{Blockchain Technology}
Blockchain is the technology behind Bitcoin~\cite{Satoshi:bitcoin} and subsequent cryptocurrencies. 
Besides the huge impact it had on the financial sectors, blockchain is considered the next generation application building technology because it provides two core elements for realising decentralisation: (i) a distributed ledger, and (ii) a decentralised "compute" infrastructure.

The underlying distributed ledger of a blockchain gives transaction storage and verification services, without relying on any central trusted authority \cite{scheuermann2015iacr}. Trust in permissionless blockchains is achieved via game theoretic incentives~\cite{Satoshi:bitcoin} to maintain a majority of honest nodes. 
In permissioned blockchains, trust is preserved by strictly managing and verifying the real-world identities of the nodes. The "compute" infrastructures of blockchain platforms refer to on-chain programmability (i.e. smart contracts)~\cite{Omohundro:2014}. Smart contracts enable complex on-chain business logic such as triggers, conditions, etc.

\begin{figure}[tpb]
	\centering
	\includegraphics[width=0.4\columnwidth]{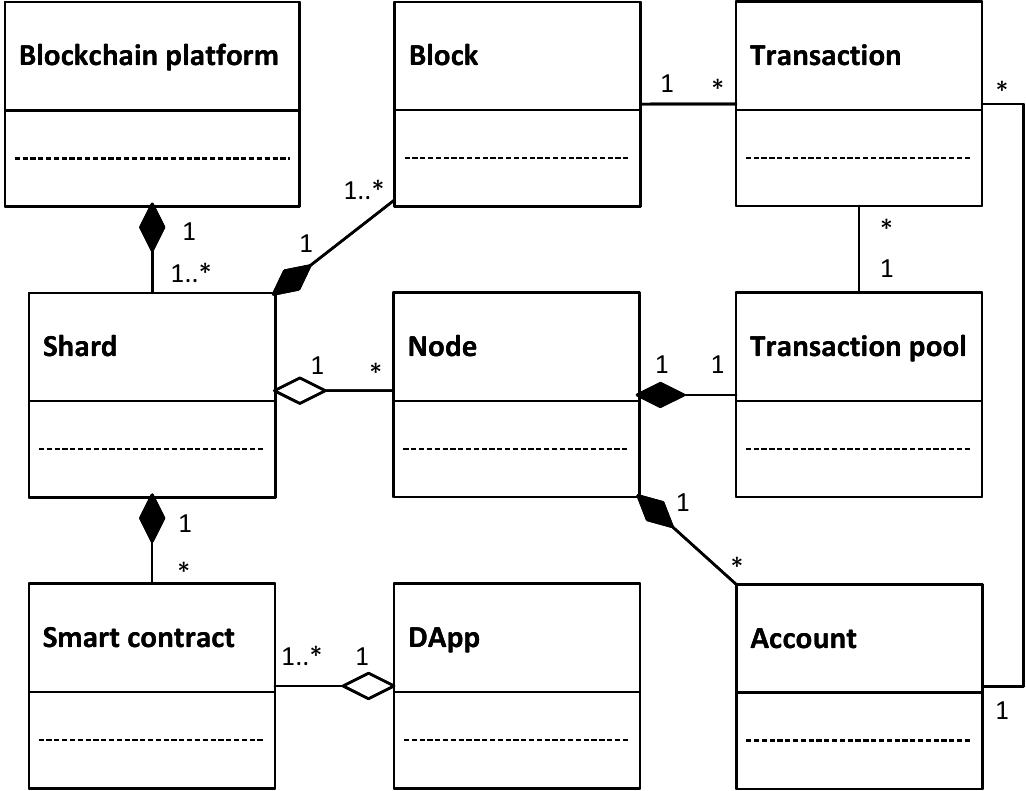}
	\caption{A conceptual representation of blockchain platforms.}
	\label{fig:component}
\end{figure}

Fig.~\ref{fig:component} illustrates a conceptual representation of the structure of a blockchain platform. A blockchain platform may consist of multiple blockchain shards. A shard may include multiple nodes. Every node holds a local replica of the shard. The data structure of blockchain is a list of identifiable blocks. All blocks (except the genesis block) are linked to the previous one and hence form a chain. Blocks are containers for transactions, which are identifiable packages carrying the changing states of on-chain data. Blockchain nodes provide interfaces for usage. Users can register multiple blockchain accounts via a blockchain node. Transaction sources are identified through users' accounts. A blockchain node also maintains a transaction pool (local memory pool), which can temporarily store the submitted transactions before they are included in a block. A node can generate candidate blocks from the pool. When a chosen candidate block is validated by the node, it is appended to the blockchain. Smart contracts can be deployed on blockchain. A decentralised application (DApp) may consist of multiple smart contracts.

\subsection{Blockchain Governance}
In our previous work, we performed a systematic literature review~\cite{liu2021systematic} to understand state-of-the-art of blockchain governance. Based on the review results, we proposed a governance framework consisting of six high-level principles~\cite{liu2021defining}. 

Essentially, blockchain practitioners apply governance for block-chain to address software qualities such as adaptability, upgradability, security, privacy, etc. These attributes can eventually improve the trustworthiness of a blockchain platform. Blockchain governance can be divided into four layers regarding where the governance methods are enforced, including blockchain platform itself, on-chain data, blockchain-based applications, and real-world collaborations in its off-chain community.

The off-chain community is comprised of different stakeholders. Typically, it includes the project team, node operators, users, application providers, and regulators. Specifically, the blockchain project team is responsible for both technical support and making any necessary real-world arrangements (e.g., maintaining formal communication channels, and interacting with other stakeholders). Node operators should hold nodes locally, which maintain all historical ledger contents of the blockchain. Users can submit transactions to use blockchain services. Application providers can choose a suitable blockchain platform to adopt in their existing workflows. Finally, regulators could be government representatives or third-party auditors, who ensure the compliance of blockchain with laws, regulations and ethical principles.

Blockchain governance highlights the allocation and execution of incentives, decision rights, and accountability over these stakeholders, to reach agreement on governance decisions. Governance methods are implemented in two ways: (i) process-driven mechanisms where the methods are realised through blockchain development process guided by a set of governance meta-rules, and (ii) product-driven mechanisms where the methods are implemented as software functionalities built into a blockchain.

\begin{figure*}[!ht]
	\centering
	\includegraphics[width=\textwidth]{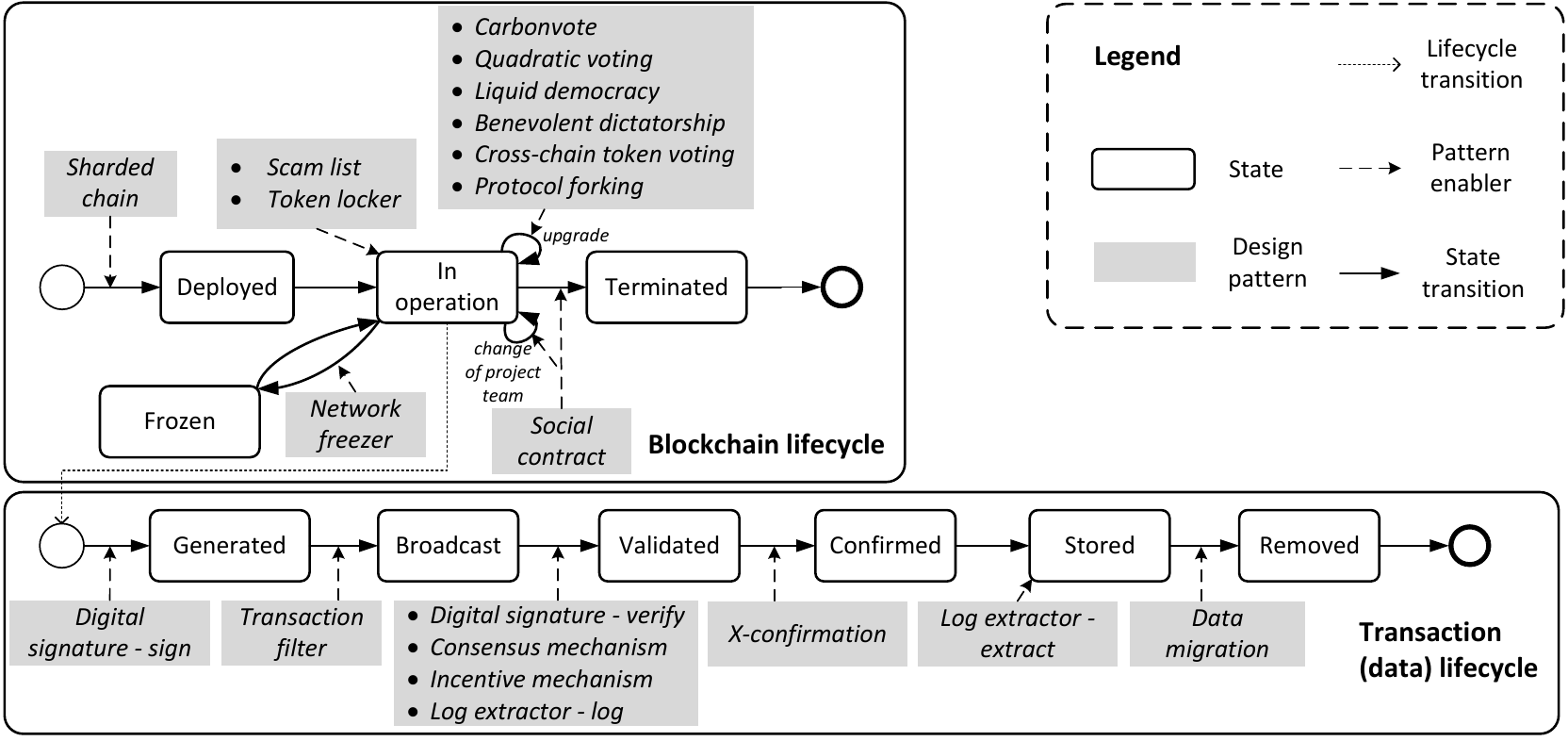}
	\caption{Lifecycle of blockchain platforms annotated with their architectural patterns.}
	\label{fig:lifecycle}
\end{figure*}

\subsection{Related Work}

In terms of software patterns in blockchain, there are some existing studies that investigated how to build blockchain-based applications using reusable solutions~\cite{xu2018pattern, oracle, eberhardt2017or}. Wohrer and Zdun~\cite{wohrer2018smart} present six patterns to address security issues in smart contract design. Zhang et al. in~\cite{zhang2017applying} share their experiences of utilising four object-oriented software patterns when designing a blockchain-based healthcare platform. There are also empirical studies identifying patterns from the perspective of blockchain ecosystem~\cite{bartoletti2017empirical, Wohrer_Zdun}. Abadin and Syed~\cite{PoW_pattern} analyse the pattern of Proof-of-Work consensus mechanism in blockchain. In our previous work, we illustrated general smart contract patterns~\cite{smartContractICBC}, self-sovereign identity patterns~\cite{SSIpattern}, and blockchain-based payment patterns~\cite{lu2021patterns}.

With regards to blockchain governance, Katina et al.~\cite{selected5} summarise seven interrelated elements in this topic. Allen and Berg~\cite{selected7} propose a descriptive framework focusing on exogenous and endogenous governance methods. Beck et al.~\cite{selected14} introduce the three dimensions of governance: decision rights, accountability, and incentives, which are adopted from IT governance. John and Pam~\cite{selected10} and Pelt et al.~\cite{selected11} both study on-chain and off-chain development processes. Howell et al.~\cite{selected15} analyse the membership and transacting relationships. Werner et al.~\cite{selected31} develop a taxonomy for platform governance in blockchain. Hofman et al.~\cite{hofman2021blockchain} propose an analytic framework regarding six aspects (i.e., why, who, when, what, where, and how). However, these studies either give an abstract discussion on the development process, or analyse coarse-grained governance layers in the blockchain ecosystem, and do not offer adequate technical implementation of governance methods. Our work presents 14 architectural patterns regarding blockchain platform lifecycle stages. The pattern language can provide a holistic guidance to practitioners on reusable governance solutions, and trade-off analysis for the design of governance-driven blockchain systems.

\begin{table*}[tbp]
\footnotesize
\centering
\caption{Pattern language overview.}
\label{overview}
\begin{tabular}{p{0.14\columnwidth}p{0.15\columnwidth}p{0.66\columnwidth}}
\toprule

{\bf Name} &
{\bf  Decentralisation level} &
\multicolumn{1}{c}{\bf Summary}\\
\midrule

\multirow{2}{0.14\columnwidth}{Sharded chain} & \multirow{2}{0.2\columnwidth}{Permissioned \& permissionless} & \multirow{2}{0.66\columnwidth}{A blockchain platform can be divided into multiple shards to process transactions in parallel.}
\\
\cmidrule(l){1-3}

%& 
\multirow{2}{0.14\columnwidth}{Scam list} & \multirow{2}{0.2\columnwidth}{Permissionless} & The blockchain addresses of the entities who are deemed malicious are labelled, and listed to warn all stakeholders.
\\
\cmidrule(l){1-3}

%& 
\multirow{2}{0.14\columnwidth}{Token locker} & \multirow{2}{0.3\columnwidth}{Permissionless} & A certain amount of blockchain tokens are locked for a specified time period, to secure the token holders' behaviour.
\\
\cmidrule(l){1-3}

%& 
\multirow{2}{0.14\columnwidth}{Network freezer} & \multirow{2}{0.2\columnwidth}{Permissioned \& permissionless} & \multirow{2}{0.66\columnwidth}{The blockchain platform is frozen that all on-chain business is suspended.}
\\
\cmidrule(l){1-3}

%& 
\multirow{2}{0.14\columnwidth}{Carbonvote} & \multirow{2}{0.2\columnwidth}{Permissionless} & Votes for improvement proposals are counted according to the tokens held by blockchain addresses.
\\
\cmidrule(l){1-3}

%& 
\multirow{2}{0.14\columnwidth}{Quadratic voting} & \multirow{2}{0.2\columnwidth}{Permissionless} & For a blockchain account, submitting $n$ votes for an improvement proposal costs $n^{2}$ tokens.
\\
\cmidrule(l){1-3}

%& 
\multirow{2}{0.14\columnwidth}{Liquid democracy} & \multirow{2}{0.3\columnwidth}{Permissionless} & A stakeholder can delegate the voting rights to other stakeholders, and revoke the delegation to directly vote for improvement proposals.
\\
\cmidrule(l){1-3}

\multirow{2}{0.12\columnwidth}{Benevolent dictatorship} & \multirow{2}{0.3\columnwidth}{Permissionless} &  A subset of developers of a blockchain platform have additional rights to finalise governance-related decisions.
\\
\cmidrule(l){1-3}

%& 
\multirow{2}{0.12\columnwidth}{Cross-chain token voting} & \multirow{2}{0.3\columnwidth}{Permissionless} & Specific token holders in a blockchain platform can vote for governance-related issues of another blockchain platform.
\\
\cmidrule(l){1-3}

%& 
\multirow{2}{0.14\columnwidth}{Protocol forking} & \multirow{2}{0.2\columnwidth}{Permissioned \& permissionless} & \multirow{2}{0.66\columnwidth}{The software upgrades of a blockchain platform are implemented as forks of the blockchain.}
\\
\cmidrule(l){1-3}

%& 
Social contract & Permissionless & A social contract is deployed to select the future maintainers of a blockchain platform.
\\
\cmidrule(l){1-3}

%& 
\multirow{2}{0.14\columnwidth}{Digital signature} & \multirow{2}{0.2\columnwidth}{Permissioned \& permissionless} & \multirow{2}{0.66\columnwidth}{Transactions are digitally signed by users, to identify the transaction sources. The digital signature can be verified by other stakeholders.}
\\
\cmidrule(l){1-3}

\multirow{2}{0.14\columnwidth}{Transaction filter} & \multirow{2}{0.2\columnwidth}{Permissioned \& permissionless} & \multirow{2}{0.66\columnwidth}{A filter can be utilised to examine the submitted transactions, to ensure the validity of transaction format/content.} 
\\
\cmidrule(l){1-3}

\multirow{2}{0.14\columnwidth}{Log extractor} & \multirow{2}{0.2\columnwidth}{Permissioned \& permissionless} & \multirow{2}{0.66\columnwidth}{Log extractor allows application providers to extract DApp usage information from blockchain.}
\\
%\cmidrule(l){1-2}

\bottomrule
\end{tabular}
\end{table*}

\begin{table}[tbp]
\footnotesize
\centering
\caption{Glossary for specific concepts.}
\label{tab:glossary}
\begin{tabular}{p{0.27\columnwidth}|p{0.64\columnwidth}}
\toprule

\multirow{1}{0.25\columnwidth}{\bf Block height} & The number of blocks preceding a specific block.\\
\cmidrule(l){1-2}

\multirow{1}{0.27\columnwidth}{\bf Blockchain address/account} & \multirow{1}{0.64\columnwidth}{Stakeholders' on-chain identifiers.}\\
\cmidrule(l){1-2}

\multirow{2}{0.25\columnwidth}{\bf Decentralised application (DApp)} & Applications built on a decentralised network. A blockchain DApp consists of multiple smart contracts and a user interface.\\
\cmidrule(l){1-2}

\multirow{2}{0.25\columnwidth}{\bf Delegated Proof of Stake (DPoS)} & A consensus mechanism where users can vote on delegates. The selected delegate can append a new block to the blockchain.\\
\cmidrule(l){1-2}

\multirow{1}{0.25\columnwidth}{\bf Security deposit} & A specific number of blockchain tokens are locked for a certain period.\\
\cmidrule(l){1-2}

\multirow{2}{0.25\columnwidth}{\bf Token} & Programmable digital assets in blockchain. In permissionless public blockchains, tokens are issued as cryptocurrencies.\\
\cmidrule(l){1-2}

\multirow{1}{0.25\columnwidth}{\bf Transaction pool} & Blockchain node's memory pool, to temporarily store the submitted transactions.\\
%\cmidrule(l){1-2}

\bottomrule
\end{tabular}
\end{table}

\section{Overview of Patterns for Blockchain Governance}
\label{methodology}

We first performed a systematic literature review (SLR). We analysed 37 primary studies with six research questions (i.e. what, why, where, when, who, how)~\cite{liu2021systematic}. The extracted answers cover the following aspects of blockchain governance: dimensions, motivations, objects, processes, stakeholders, and mechanisms. We also reviewed extant governance frameworks and standards, including IT governance~\cite{weill2004governance, cobit2012business}, data governance~\cite{ballard2014ibm, ISO38505}, OSS governance~\cite{o2007emergence, de2007governance}, and platform ecosystem governance~\cite{tiwana2010platform}. The review process helped us identify the salient characteristics of blockchain governance. We then proposed six high-level governance principles and a governance framework~\cite{liu2021defining} designed from the principles. The framework was evaluated by scrutinising the open websites and documents of five blockchain platforms (i.e., Bitcoin, Ethereum, Dash, Tezos, and Hyperledger Fabric), to understand how blockchain governance is realised in practice, and whether our principles are applied in real-world context.

The pattern language is derived from our previous literature review, and the experiences gained from developing the governance framework. We explore more extant blockchain platforms, through reading the official websites and documents available, to identify and validate the proposed patterns. Fig.~\ref{fig:lifecycle} illustrates the lifecycle of a blockchain platform. Each stage or stage transition in this lifecycle is associated with the patterns.

At the start of the lifecycle, before a blockchain platform is officially deployed, the underlying infrastructure should be decided. \textit{Sharded chain} means that a blockchain is divided into multiple shards. Each shard can individually process on-chain transactions. When a blockchain platform is in operation phase, \textit{scam list} and \textit{token locker} can restrict stakeholders' behaviour by imposing penalties. When emergencies occur, \textit{network freezer} pauses on-chain activities until the system is recovered via human interventions.

A blockchain platform needs upgrades and improvements, for instance, by fixing software bugs or adding new functionalities. A series of patterns can help finalise the improvement proposals. These include \textit{carbonvote, quadratic voting, liquid democracy, benevolent dictatorship, cross-chain token voting,} and \textit{protocol forking} for upgrade implementation. When the blockchain platform stops providing further services, a \textit{social contract} can be deployed to select or announce the candidate maintainer(s) of this blockchain.

When a blockchain transaction is generated, it must contain a \textit{digital signature} of the corresponding user. The transaction needs to be checked by \textit{transaction filter} of a node, before it is stored in the node's local transaction pool, and broadcast to other nodes. Afterwards, the \textit{digital signature} in transactions are verified by validators. Valid transactions are collected into a block. The block is then appended to blockchain according to the employed \textit{consensus mechanism}, along with \textit{incentive mechanism} to reward the block validator. A block and its contained transactions are regarded as officially recorded on-chain after \textit{X-confirmation}, which means that certain numbers of subsequent blocks are appended. Stored transactions can be reviewed and analysed via \textit{log extractor}. Finally, on-chain data can be migrated from one blockchain instance to another via \textit{data migration} patterns.

Please note that in this paper, we focus on 14 of the above patterns, as existing works already explored \textit{consensus mechanism}~\cite{consensus_survey, PoW_pattern}, \textit{incentive allocation}~\cite{liu2021defining}, \textit{X-confirmation}~\cite{xu2018pattern}, and \textit{data migration}~\cite{data_migration}. Table~\ref{overview} offers an overview of the 14 patterns.

\section{Pattern Language for Blockchain Governance}
\label{patterns}

To describe each pattern, we adopt the extended pattern form introduced in \cite{patternLanguage}. It includes the pattern name, a short summary, usage context, problem statement, discussion of forces, the solution and its consequences, and real-world known uses of the pattern. Please note that applying the patterns may bring consequences additional to the forces. Table~\ref{tab:glossary} provides a glossary of specific concepts to help understand the pattern language.

\subsection{Sharded Chain}

\vspace{0.5em}\noindent \textbf{Summary:} A blockchain platform can be divided into multiple shards to process transactions in parallel.

\vspace{0.5em}\noindent \textbf{Context:} Normally, there is only one blockchain instance as the main network in a blockchain platform. All stakeholders reference the same blockchain instance. All data are recorded in this instance.

\vspace{0.5em}\noindent \textbf{Problem:} How can node operators handle a large number of transactions and be able to maintain an ever-growing blockchain?

\vspace{0.5em}\noindent \textbf{Forces:} 

\begin{itemize}
  \item \textit{Scalability improvement.} Blockchain's scalability should be improved. A single blockchain instance limits the scalability of the overall blockchain platform. Only one block is appended to the blockchain each round, based on predefined block size and interval.

  \item \textit{Cost reduction.} The participating nodes need to maintain a replica of all historical ledger contents. The cost of maintaining a blockchain node should be reduced.

\end{itemize}

\begin{figure}[!ht]
	\centering
	\includegraphics[width=0.3\columnwidth]{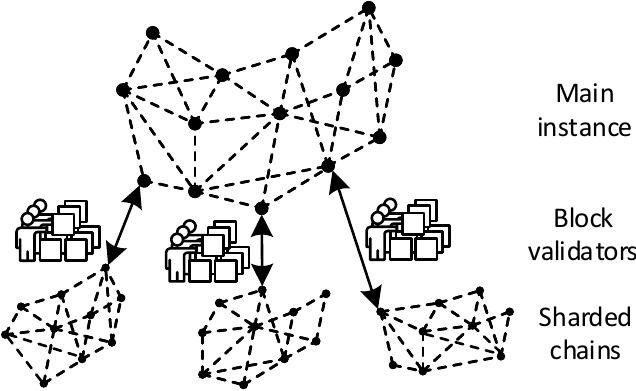}
	\caption{Sharded chain.}
	\label{pic:shard_chain}
\end{figure}

\vspace{0.5em}\noindent \textbf{Solution:} Fig.~\ref{pic:shard_chain} is a graphical representation of \textit{sharded chain}. When developing a blockchain platform, the project team needs to consider the decentralisation level, and determine the underlying infrastructure. \textit{Sharded chain} refers to a blockchain platform consisting of multiple blockchain instances as shards, where transaction validation, data storage, and on-chain computation are processed in parallel. A node operator can maintain full node(s) for either a single shard or multiple shards locally. Block validators are randomly assigned to different shards to generate and append new blocks. In addition, the project team can implement main instance(s) acting as the coordinator(s) between shards. The main instance(s) need to record the block information of all shards.

\vspace{0.5em}\noindent \textbf{Consequences:} 

Benefits:
\begin{itemize}
  \item \textit{Scalability.} The overall scalability is improved, as multiple shards can process transactions in parallel.
  
  \item \textit{Cost.} For a single node operator, he/she only needs to maintain a node for one shard, instead of the historical transactions for the whole blockchain. This can reduce the management overhead and cost of storage.
  
  \item \textit{Interoperability.} The interoperability between different shards is achieved by cross-shard transactions, which should be stored in both original and target shards.
  
\end{itemize}

Drawbacks:
\begin{itemize}
  \item \textit{Security.} A shard may be compromised when multiple malicious block validators are assigned to this shard. This is known as 1\% attack~\cite{single_shard_attack}.

  \item \textit{Cost.} Sharded chains may increase the cost of deployment, communication, and maintenance for cross-shard nodes. Cross-shard nodes are responsible to maintain the ledger contents of multiple shards. There may be redundant storage in cross-shard nodes, since cross-shard transactions need to be recorded by all relevant shards.

\end{itemize}

\vspace{0.5em}\noindent \textbf{Known uses:}  
 \begin{itemize}
   \item \textit{Ethereum}\footnote{\url{https://www.ethereum.org/}\label{ethereum}}. Ethereum is undergoing an update where 64 shards will be introduced to improve scalability.
   
   \item \textit{NEAR}\footnote{\url{https://near.org/}}. NEAR aims to support cross-shard transactions. Block validators only need to focus on a single shard. Sharding can lower the hardware requirements of nodes.
   
   \item \textit{Polkadot}\footnote{\url{https://polkadot.network/}}. Polkadot deploys a "relay chain" as the main blockchain instance, while the shards are called "parachains".

 \end{itemize}

\vspace{0.5em}\noindent \textbf{Related patterns:} 

\begin{itemize}
    \item \textit{Consensus mechanism}~\cite{consensus_survey, PoW_pattern}. Different consensus mechanisms can be employed in the multiple shards.
\end{itemize}

\subsection{Scam List}

\vspace{0.5em}\noindent \textbf{Summary:} The blockchain addresses of the entities who are deemed malicious are labelled, and listed to warn all stakeholders.

\vspace{0.5em}\noindent \textbf{Context:} Blockchain provides a peer-to-peer platform, where any two stakeholders can directly interact with each other. There is no authority to handle on-chain conflicts and even malicious activities.

\vspace{0.5em}\noindent \textbf{Problem:} How can a project team exclude malicious entities with suspicious on-chain activities from the platform?

\vspace{0.5em}\noindent \textbf{Forces:} 

\begin{itemize}
  \item \textit{Security.} Stakeholders should be protected from on-chain frauds.

  \item \textit{Accountability guarantee.} Attackers should be accountable for their malicious behaviours.
\end{itemize}

\vspace{0.5em}\noindent \textbf{Solution:} %Fig.~\ref{pic:scam_list} is a graphical representation of \textit{scam list}. 
The project team of a blockchain platform can record the suspicious or verified malicious blockchain addresses and DApps. This record can be either an on-chain smart contract, or an off-chain post on the platform's official website. The listed blockchain addresses are then labelled for being risky. Therefore, other stakeholders can refer to this scam list, to avoid interacting with the recorded blockchain addresses.

\vspace{0.5em}\noindent \textbf{Consequences:} 

Benefits:
\begin{itemize}
  \item \textit{Security.} Stakeholders are warned that interacting with certain blockchain addresses may jeopardise their business profits due to potential fraud.
  
  \item \textit{Accountability.} Blockchain addresses involved in on-chain scams are recorded.
\end{itemize}

Drawbacks:
\begin{itemize}
  \item \textit{Security.} \textit{Scam list} is an ex-post measure where malicious blockchain addresses are reported and recorded after the scams. The interests of certain stakeholders may have already been harmed.
  
  \item \textit{Accountability.} (i) Scammers may relate themselves to other ordinary blockchain addresses via a dust attack~\cite{BRADBURY20135}. Originally, dust attacks refer to sending a large numbers of transactions with a small cost, to affect the performance of blockchain. In the case of scam detection, dust attacks can connect malicious addresses with other ordinary addresses, which increases the difficulty of detecting the malicious addresses. (ii) In permissionless blockchain platforms, it is hard to identify the malicious stakeholders in real-world due to the inherent anonymity.
\end{itemize}

\vspace{0.5em}\noindent \textbf{Known uses:}  
 \begin{itemize}
   \item \textit{Ethereum}\textsuperscript{\ref{ethereum}}. Ethereum discusses common scams to prevent serious risks in online posts.
 
   \item \textit{Dash}\footnote{\url{https://www.dash.org/}\label{dash}}. Dash maintains a page in its official online documents, providing safety guidelines and listing the known scams in Dash blockchain.
   
   \item \textit{Tezos\footnote{\url{https://tezos.com/}\label{tezos}}.} The Tezos Foundation monitors the Tezos blockchain, and tracks malicious activities. The known scams are listed on their official website.
 \end{itemize}

\vspace{0.5em}\noindent \textbf{Related patterns:} 

\begin{itemize}
    \item \textit{Token locker.} In permissionless blockchains, the tokens of a black-listed blockchain addresses can be permanently locked as a penalty.
    
    \item \textit{Log extractor.} Suspicious blockchain addresses can be tagged and alarmed to all stakeholders after analysing extracted logs.
\end{itemize}

\subsection{Token Locker}

\vspace{0.5em}\noindent \textbf{Summary:} A certain amount of blockchain tokens are locked for a specified time period, to secure the token holders' behaviour.

\vspace{0.5em}\noindent \textbf{Context:} Permissionless blockchain platforms issue tokens which are valuable digital assets that enable token holders to participate in important operational matters (e.g., voting).

\vspace{0.5em}\noindent \textbf{Problem:} How can a project team restrict token holders' behaviour in such a way that they are prevented from carrying out malicious attacks?

\vspace{0.5em}\noindent \textbf{Forces:} 

\begin{itemize}
  \item \textit{Security.} A proper solution is required to impose restrictions on token holders' behaviour.
  
  \item \textit{Flexible decision rights.} Token holders may be interested in certain decision-makings, and request for relevant decision rights.
  
  \item \textit{Fast monitoring.} Operations of token holders should be monitored, and be responded to in a short amount of time.
\end{itemize}

\vspace{0.5em}\noindent \textbf{Solution:}  
In permissionless blockchains, a token holder needs to lock certain amounts of tokens as a security deposit when conducting specific activities, for instance, participating in a governance decision-making process. The locked tokens cannot be used for other purposes, until the decision is finalised. Otherwise, the holder will lose the relevant decision rights. Further, if fraud is detected in a decision-making process, the accountable blockchain addresses will be banned for any on-chain transaction, and their tokens will be locked. In more serious cases, the locked tokens will be burned (aka. destroyed), and cannot be restored.

\vspace{0.5em}\noindent \textbf{Consequences:} 

Benefits:
\begin{itemize}
  \item \textit{Security.} Malicious operations will lead to the loss of real money, which discourages token holders from misbehaving when they are exercising their rights with tokens.
  
  \item \textit{Flexibility.} Decision rights can be flexibly granted to token holders who own and lock the required number of tokens as a deposit.
  
  \item \textit{Fast monitoring.} The lock, return, and destruction of tokens can be immediately processed according to token holders' behaviour.
\end{itemize}

Drawbacks:
\begin{itemize}
  \item \textit{Security.} Malicious attackers may not mind the loss of tokens. Their purpose is to disturb the blockchain.

  \item \textit{Centralisation.} Holders possessing more tokens can have more decision rights. 

  \item \textit{Deflation.} Commonly, the maximum supply of tokens in permissionless blockchains is settled at the beginning of the design and development. Burning tokens will reduce the circulating supply of tokens, and cause deflation.
\end{itemize}

\vspace{0.5em}\noindent \textbf{Known uses:}  
 \begin{itemize}
   \item \textit{Dash}\textsuperscript{\ref{dash}}. Node operators need to possess 1,000 Dash tokens to become a Masternode.
   
   \item \textit{Tezos}\textsuperscript{\ref{tezos}}. Validating a block requires a certain amount of Tezos tokens, which can be retrieved after a predefined cycle (e.g., one year).
   
   \item Baudlet et al. \cite{selected2} propose a framework, in which node operators can run Masternodes by locking tokens for a variable time period.
 \end{itemize}

\vspace{0.5em}\noindent \textbf{Related patterns:} 

\begin{itemize}
    \item \textit{Scam list.} The tokens of recorded addresses in a \textit{scam list} can be locked and burned as a penalty.

    \item \textit{Carbonvote.} In a \textit{carbonvote} process, a token holder needs to keep their tokens in a blockchain address, which will be counted as votes to finalise an improvement proposal.
    
    \item \textit{Consensus mechanism}~\cite{consensus_survey}. In certain \textit{consensus mechanisms} (e.g., Proof-of-Stake), node operators need to possess a certain amount of tokens, to participate in the election of block validators.
\end{itemize}

\subsection{Network Freezer}

\vspace{0.5em}\noindent \textbf{Summary:} The blockchain platform is frozen that all on-chain business is suspended.

\vspace{0.5em}\noindent \textbf{Context:} Emergency cases may risk the normal operation of blockchain.

\vspace{0.5em}\noindent \textbf{Problem:} How can a project team handle emergencies, and prevent the emergency situations from affecting the blockchain?

\vspace{0.5em}\noindent \textbf{Forces:} 

\begin{itemize}
  \item \textit{Security.} Blockchain and deployed smart contracts should be protected from malicious transactions.

  \item \textit{Fast monitoring.} Malicious transactions should be immediately stopped.

\end{itemize}

\begin{figure}[!ht]
	\centering
	\subfigure[Terminating smart contract.]{
	\centering
    \includegraphics[width=0.3\columnwidth]{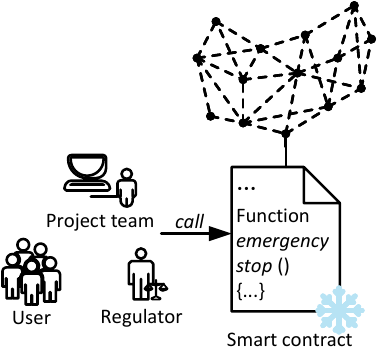}
    }
	\subfigure[Blocking Internet data traffic.]{
	\centering
    \includegraphics[width=0.19\columnwidth]{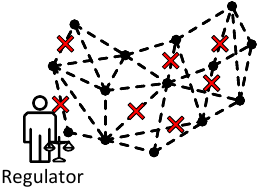}
    }
    \caption{Network freezer.}
	\label{pic:network_freezer}
\end{figure}

\vspace{0.5em}\noindent \textbf{Solution:}  Fig.~\ref{pic:network_freezer} is a graphical representation of \textit{network freezer}. Applying this pattern requires the interference of the project team, or even regulators to suspend all business in the blockchain platform. This can be accomplished in two manners. (i) As shown in Fig.~\ref{pic:network_freezer} (a), each deployed smart contract should implement a function for "emergency stop", which enables relevant stakeholders (e.g., the contract developer, regulators, and project team) to pause or terminate the operation of this smart contract. (ii) Regulators can block the data traffic of this blockchain platform via local Internet providers, as illustrated in Fig.~\ref{pic:network_freezer} (b). Nodes cannot broadcast transactions or blocks to each other, therefore, the whole blockchain is frozen.

\vspace{0.5em}\noindent \textbf{Consequences:} 

Benefits:
\begin{itemize}
  \item \textit{Security.} Attacks towards a blockchain platform are suspended, as well as all other on-chain businesses. This allows stakeholders to have enough time to find the appropriate solutions.
  
  \item \textit{Fast monitoring.} All on-chain businesses can be paused in a short amount of time.
\end{itemize}

Drawbacks:
\begin{itemize}
  \item \textit{Centralisation.} This pattern defines privileged stakeholders, who are able to freeze the blockchain.
  
  \item \textit{Security.} If privileged stakeholders are compromised, malicious stakeholders may launch denial of service attacks.
  
  \item \textit{Cost.} All other valid transactions are postponed, until the next human intervention to reactivate the blockchain.
\end{itemize}

\vspace{0.5em}\noindent \textbf{Known uses:}  
 \begin{itemize}
   \item \textit{Hyperledger Fabric}\footnote{\url{https://www.hyperledger.org/use/fabric}}. Smart contracts can be manually stopped, and removed from nodes.
 
   \item Developing plausible smart contracts and self-destruct functions in Ethereum are analysed in \cite{wohrer2018smart}.
   
   \item Enforcing network restrictions is discussed in~\cite{selected16, selected25}. This solution is viewed as a tough measure for extreme cases.
 \end{itemize}

\vspace{0.5em}\noindent \textbf{Related patterns:} 

\begin{itemize}
    \item \textit{Network freezer} is not closely related to specific patterns. In general, freezing the network affects all on-chain transactions, including the application of other patterns.

\end{itemize}

\subsection{Carbonvote}

\vspace{0.5em}\noindent \textbf{Summary:} Votes for improvement proposals are counted according to the tokens held by blockchain addresses.

\vspace{0.5em}\noindent \textbf{Context:} In permissionless blockchains, token holders need to vote for improvement proposals, by sending transactions from their blockchain addresses, to determine the upgrade proposals of the blockchain platform.

\vspace{0.5em}\noindent \textbf{Problem:} Every blockchain user can possess multiple blockchain addresses to vote. Therefore, calculating the number of voting addresses is unreliable. How can a project team cast a viable vote to finalise improvement proposals, when counting heads (i.e. voted addresses) is not applicable?

\vspace{0.5em}\noindent \textbf{Forces:} 

\begin{itemize}
  \item \textit{Security.} Improvement proposal votes should be protected from Sybil attack. In a Sybil attack, malicious attackers may register multiple blockchain addresses and compromise the vote.

  \item \textit{Fairness guarantee.} Improvement proposal votes should be fair that token holders are granted decision rights according to their stakes in the blockchain. In permissionless blockchains, owning more tokens is considered as having more stakes in the blockchain.

  \item \textit{Efficiency improvement.} Differentiating valid and invalid votes may be tedious. Counting votes should be efficient to determine whether to accept or reject an improvement proposal.
\end{itemize}

\begin{figure}[!ht]
	\centering
	\includegraphics[width=0.43\columnwidth]{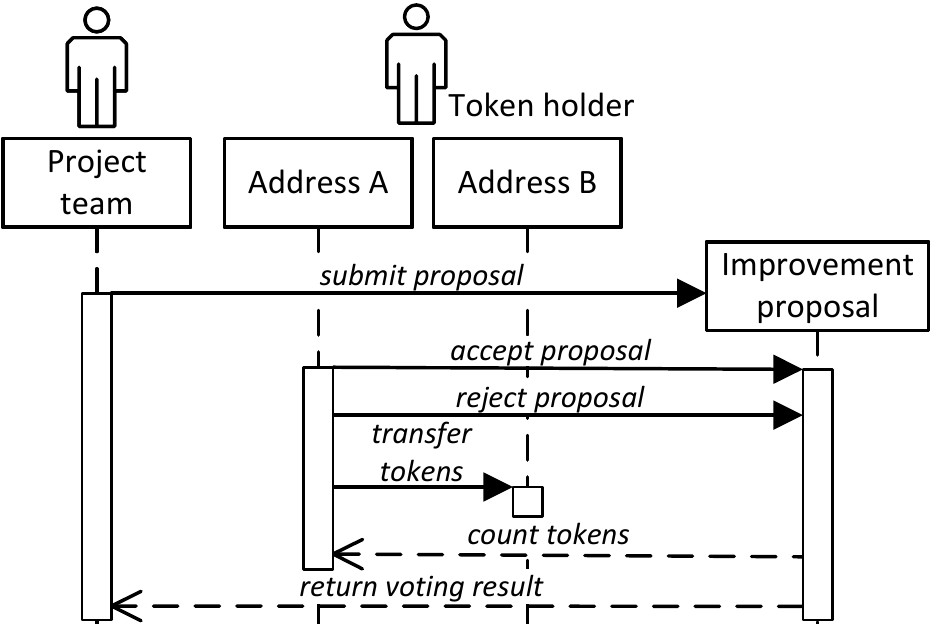}
	\caption{Carbonvote.}
	\label{pic:Carbonvote}
\end{figure}

\vspace{0.5em}\noindent \textbf{Solution:} Fig.~\ref{pic:Carbonvote} illustrates the sequence of \textit{carbonvote} to address the above challenges. First, the project team submits an improvement proposal, and starts a voting process. Let's assume that a token holder has two blockchain addresses. The token holder first sends a transaction to vote for acceptance via address A. The blockchain platform will monitor the tokens possessed by address A in real-time. The token holder can change his/her vote to another choice (rejection in this example), by sending an additional transaction. Besides, he/she can transfer all tokens in address A to another address owned by himself/herself (address B in this example). Therefore, address A will have no tokens counted as votes. Finally, the choice with the most tokens will be accepted as the voting result.

\vspace{0.5em}\noindent \textbf{Consequences:} 

Benefits:
\begin{itemize}
  \item \textit{Security.} \textit{Carbonvote} can prevent Sybil attack as additional blockchain addresses won't be counted as votes. Votes are cast according to the number of tokens instead of accounts.
  
  \item \textit{Fairness.} Votes are allocated to users according to the tokens they possessed. Users with more tokens are considered to have more contributions and stakes to the operation of a blockchain. Therefore, these users have more decision rights (in the form of votes).
  
  \item \textit{Efficiency.} The voting result can be efficiently calculated, as the blockchain monitors the tokens owned by voted addresses in real-time.
\end{itemize}

Drawbacks:
\begin{itemize}
  \item \textit{Security.} The carbonvote process may be compromised by flash loan attacks, where an attacker can borrow a large amount of blockchain tokens for voting, and return the tokens when the vote ends.

  \item \textit{Centralisation.} The majority of votes may be controlled by users who possess the most tokens. This may cause plutocracy in permissionless blockchains where wealthy parties are centred with the most decision rights.
  
  \item \textit{Cost.} Although voting does not consume tokens, users still need to pay fees for sending transactions to trigger the votes.
\end{itemize}

\vspace{0.5em}\noindent \textbf{Known uses:}  
 \begin{itemize}
   \item \textit{DAO hard fork}\footnote{\url{https://ethereum.org/en/governance/}}. The Ethereum DAO hard fork was finalised via carbonvote. Over 85\% votes agreed to fork.
   
   \item \textit{EIP\#186\footnote{\url{https://github.com/ethereum/EIPs/issues/186}}.} The acceptance of EIP\#186 in Ethereum is decided via carbonvote.
 \end{itemize}
 
\vspace{0.5em}\noindent \textbf{Related patterns:} 

\begin{itemize}

    \item \textit{Token locker.} In a \textit{carbonvote} process, token holders need to keep their tokens in a blockchain address as votes, which is similar to locking tokens.
    
    \item \textit{Cross-chain token voting.} \textit{Carbonvote} can be integrated with \textit{cross-chain token voting} to finalise on-chain decision-makings.
    
\end{itemize}

\subsection{Quadratic Voting}

\vspace{0.5em}\noindent \textbf{Summary:} For a blockchain account, submitting $n$ votes for an improvement proposal costs $n^{2}$ tokens.

\vspace{0.5em}\noindent \textbf{Context:} In permissionless blockchains, token holders need to vote for improvement proposals, by sending transactions with tokens from their blockchain accounts, which determines if a proposed upgrade/change is to be implemented.

\vspace{0.5em}\noindent \textbf{Problem:} How can a project team preserve a fair voting process where token holders can also express the strength of their preferences (i.e., highly agreeing, highly disagreeing)?

\vspace{0.5em}\noindent \textbf{Forces:} 

\begin{itemize}
  \item \textit{Security.} Improvement proposal votes should be protected from Sybil attack.
  
  \item \textit{Preference expression.} Votes should express token holders' stregnth of their preferecnes towards the choice of accepting or not accepting an improvement proposal. The scheme of "one person one vote" does not accurately capture the preferences. For instnace, there may be many token holders expressing yes with low preference strength, and few token holders expressing no with high preference strength.

  \item \textit{Fairness.} In the scheme of "one token one vote", the one with the most tokens can dominate the voting process, which is unfair to other token holders. The voting should be as fair as possible so all users can have some influence in decisions.

\end{itemize}

\begin{figure}[!ht]
	\centering
	\includegraphics[width=0.43\columnwidth]{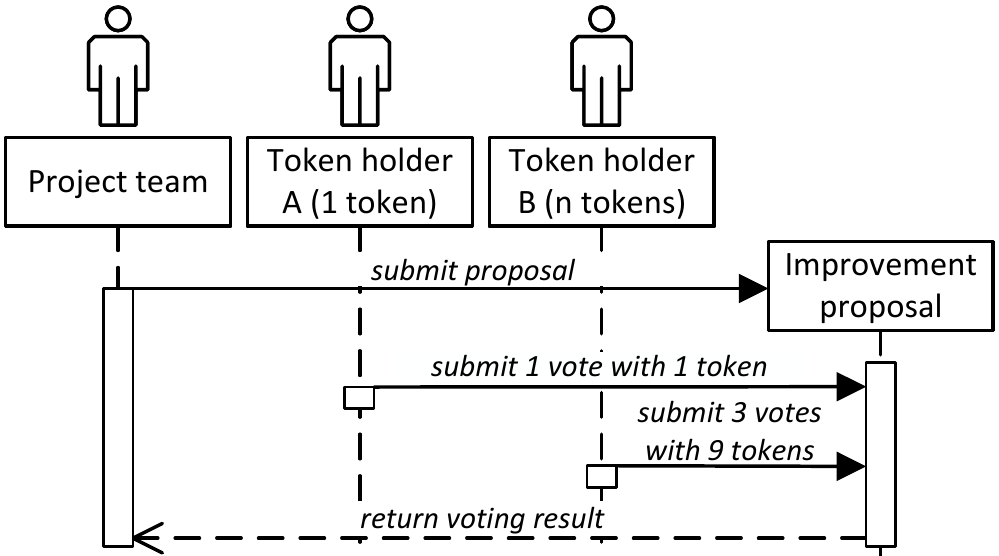}
	\caption{Quadratic voting.}
	\label{pic:quadratic_voting}
\end{figure}

\vspace{0.5em}\noindent \textbf{Solution:} As depicted in Fig.~\ref{pic:quadratic_voting}, \textit{quadratic voting} can help capture the preferences of token holders, while preventing wealthier token holders from manipulating the voting process. After the project team submits an improvement proposal, token holders can vote for its acceptance/rejection. In this voting scheme, tokens are consumed as funds for the improvement proposal. For any blockchain account, the cost is calculated as follows:

\centerline{number of tokens = (number of votes)$^{2}$}

\noindent
As shown in the figure, token holder A only has 1 token, hence, it can only submit one vote for the improvement proposal. Token holder B has $n$ tokens, and it decides to submit 3 votes for the proposal. The votes will cost it 9 tokens. The choice with the most votes will be finalised for further upgrade implementation.

\vspace{0.5em}\noindent \textbf{Consequences:} 

Benefits:
\begin{itemize}
  \item \textit{Security.} Sybil attack can be prevented, as token holders need to consume tokens to submit their votes.
  
  \item \textit{Preference.} Each additional vote requires a quadratic increase in token consumption, which can indicate how strongly token holders prefer their decisions.
  
  \item \textit{Fairness.} The high cost of additional votes can reduce the impact from wealthier token holders, to provide a fair voting process for other token holders.
\end{itemize}

Drawbacks:
\begin{itemize}
  %\item \textit{Security.} A malicious attacker may register multiple blockchain accounts, and distribute 1 token to each account. In this way, the attacker does not have to pay the quadratically-increased number of tokens for additional votes. Such operations may include cumbersome manual operations, and transaction fees for token distribution.

  \item \textit{Fairness.} Although fairness has been improved, richer participants still have more power than poorer ones.
  
  \item \textit{Cost.} Finalising improvement proposals costs resources of token holders (e.g., real money in permissionless blockchains).
\end{itemize}

\vspace{0.5em}\noindent \textbf{Known uses:}  
 \begin{itemize}
   \item \textit{Gitcoin}\footnote{\url{https://gitcoin.co/blog/}}. Gitcoin enables quadratic funding for public goods in the Ethereum ecosystem. Quadratic funding is a variant of \textit{quadratic voting} in the use case of individual provision of public goods. In quadratic funding, the funding of a project is calculated as: (i) the square root of each funder's contribution, (ii) sum up the square roots, and (iii) calculate the square of the sum~\cite{buterin2018liberal}.
   
   \item \textit{Kickflow}\footnote{\url{https://kickflow.io/}}. The Kickflow community can support projects on Tezos through quadratic funding.
   
   \item \textit{Pomelo}\footnote{\url{https://pomelo.io/grants}}. Pomelo is an open-source crowdfunding platform based on EOS. It supports quadratic funding.
 \end{itemize}

\vspace{0.5em}\noindent \textbf{Related patterns:} 

\begin{itemize}
    \item \textit{Cross-chain token voting.} \textit{Quadratic voting} can be integrated with \textit{cross-chain token voting} to finalise on-chain decision-makings.
\end{itemize}

\subsection{Liquid Democracy}

\vspace{0.5em}\noindent \textbf{Summary:} A stakeholder can delegate the voting rights to other stakeholders, and revoke the delegation to directly vote for improvement proposals.

\vspace{0.5em}\noindent \textbf{Context:} In permissionless blockchains, stakeholders need to vote for improvement proposals to determine the upgrade proposals of the blockchain platform.

\vspace{0.5em}\noindent \textbf{Problem:} If a stakeholder does not have the technical expertise of blockchain technology, how can he/she make a responsible decision on proposal voting?

\vspace{0.5em}\noindent \textbf{Forces:} 

\begin{itemize}
  \item \textit{Participation rate improvement.} The participation rate for finalising an improvement proposal should be improved. Ordinary stakeholders may be reluctant to vote for improvement proposals, due to the lack of professional knowledge of blockchain technology.

  \item \textit{Effectiveness improvement.} Improvement proposals should be finalised effectively. Stakeholders may vote when they do not fully understand the intentions of improvement proposals. This will affect the effectiveness of final results.

  \item \textit{Flexibility improvement.} The voting rights should be flexible. An inflexible voting process may cause the waste of votes that that many stakeholders do not vote.
  %A stakeholder should be able to ask others to vote on behalf of himself/herself.
\end{itemize}

\begin{figure}[!ht]
	\centering
	\includegraphics[width=0.5\columnwidth]{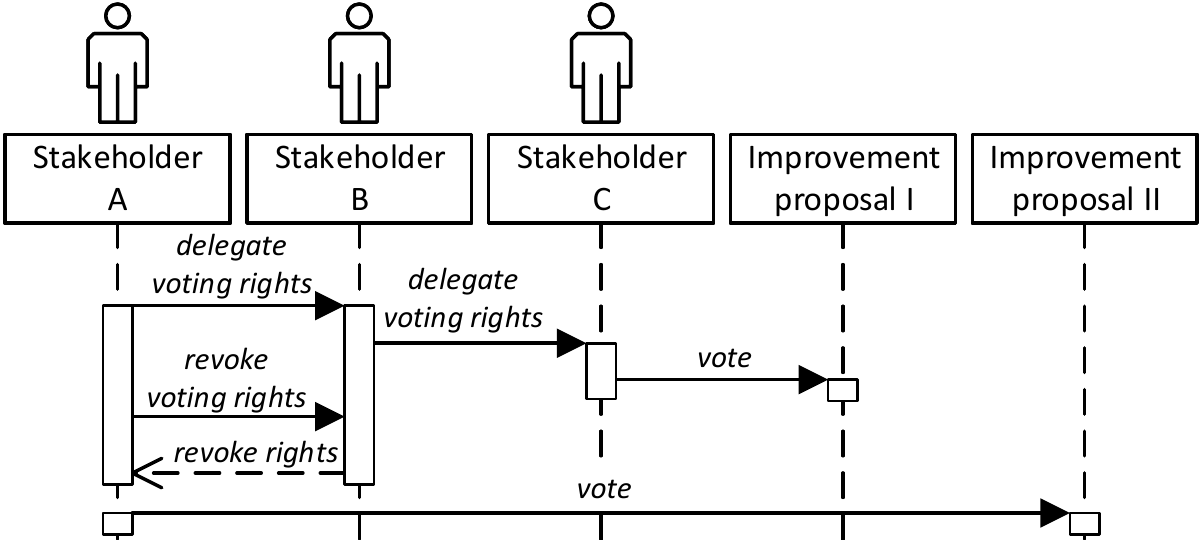}
	\caption{Liquid democracy.}
	\label{pic:liquid_democracy}
\end{figure}

\vspace{0.5em}\noindent \textbf{Solution:} As demonstrated in Fig.~\ref{pic:liquid_democracy}, \textit{liquid democracy} allows stakeholders to transfer their voting rights to trusted stakeholders. Specifically, stakeholder A first delegates stakeholder B the voting rights for improvement proposals. Subsequently, stakeholder B specifies stakeholder C as the proxy for voting. In this regard, stakeholder C can submit three votes for improvement proposal I including his/her own vote. For proposal II, stakeholder A has adequate knowledge, and decides to vote on his/her own. He/she can revoke the voting rights from stakeholder B, and then make a direct vote by himself/herself.

\vspace{0.5em}\noindent \textbf{Consequences:} 

Benefits:
\begin{itemize}
  \item \textit{Participation rate.} An entity can be delegated the voting rights from multiple stakeholders, who do not intend to vote. The vote from this entity embodies the decisions of multiple stakeholders, hence, improves the participation rate.

  \item \textit{Effectiveness.} The voter is considered an expert in blockchain technology, hence he/she is delegated to vote. He/she can make effective decisions on blockchain upgrade issues to a greater extent.
    
  \item \textit{Flexibility.} \textit{Liquid democracy} enables flexible delegation and revocation of decision rights among different stakeholders.
\end{itemize}

Drawbacks: 
\begin{itemize}

   \item \textit{Accountability.} \textit{Liquid democracy} complicates the accountability of decision-making process. All delegation should be traceable and identifiable.
   
   \item \textit{Centralisation.} \textit{Liquid Democracy} may cause centralisation during a voting process, if multiple stakeholders delegate the same voter.
\end{itemize}

\vspace{0.5em}\noindent \textbf{Known uses:}  
 \begin{itemize}
   \item \textit{DFINITY}\footnote{\url{https://dfinity.org/}}. In DFINITY, \textit{liquid democracy} is facilitated that a neuron holder can assign other neuron holders as delegates. Note that a neuron is converted from a certain number of locked tokens. When the majority of followee neurons make a particular vote, the governance canister (i.e., a special kind of smart contract for governance issues) of follower neurons will automatically record the corresponding vote.
 
   \item \textit{BitShares}\footnote{\url{https://bitshares.build/}}. BitShares deploys DPoS as the consensus mechanism.
   
   \item \textit{Compound}\footnote{\url{https://compound.finance/}}. Compound allows token owners to delegate their voting rights to other stakeholders.
 \end{itemize}

\vspace{0.5em}\noindent \textbf{Related patterns:} 

\begin{itemize}
   
    \item \textit{Benevolent dictatorship.} \textit{Benevolent dictatorship} is similar to \textit{liquid democracy} when other stakeholders delegate their decision rights to the core developers.
    
    \item \textit{Consensus mechanism}~\cite{consensus_survey}. Delegated Proof-of-Stake (DPoS) can be considered a variant of liquid democracy. Token holders are able to vote for delegates, who then can validate and append new blocks.
\end{itemize}

\subsection{Benevolent Dictatorship}

\vspace{0.5em}\noindent \textbf{Summary:} A subset of developers of a blockchain platform have additional rights to finalise governance-related decisions.

\vspace{0.5em}\noindent \textbf{Context:} There are few stakeholders at the initial phases of a blockchain lifecycle.

\vspace{0.5em}\noindent \textbf{Problem:} How governance decisions are finalised with a small community?

\vspace{0.5em}\noindent \textbf{Forces:} 

\begin{itemize}
  \item \textit{Upgradability guarantee.} Blockchain needs upgrades to fix software bugs, and implement new functionalities.

  \item \textit{Effectiveness improvement.} Governance decisions should be finalised effectively. When the platform is newly deployed, most stakeholders may not have an adequate technical background of blockchain.

  \item \textit{Quick decision.} Certain governance decisions need to be finalised within a short time.
\end{itemize}

\vspace{0.5em}\noindent \textbf{Solution:} 
\textit{Benevolent dictatorship} refers to the situation where the founder or core developers of a blockchain platform are granted great power. They can persuade others on significant governance issues. Other stakeholders trust their decisions for two main reasons: (i) the finalised governance choices can improve the blockchain, based on the expertise of technical meritocracy; (ii) the decisions are for the collective benefits of whole community, instead of personal interests of a small subset of stakeholders, as the blockchain platform needs to attract more users to survive (especially for permissionless ones). This pattern will remain in permissioned blockchains, as there are always authorities finalising governance decisions. While in permissionless blockchain platforms, it will be gradually replaced by more democratic decision-making patterns when the community becomes mature.

\vspace{0.5em}\noindent \textbf{Consequences:} 

Benefits:
\begin{itemize}
  \item \textit{Upgradability.} The blockchain platform is upgraded according to the decisions made by \textit{benevolent dictators}.
  
  \item \textit{Effectiveness.} \textit{Benevolent dictators} can make effective governance decisions for a blockchain platform based on their specialised knowledge.
    
  \item \textit{Quick decision.} A governance decision can be finalised within a short time, as there is usually only one or at most several \textit{benevolent dictators} in a blockchain community.
\end{itemize}

Drawbacks:
\begin{itemize}
  \item \textit{Centralisation.} Most decision rights are centred on the \textit{benevolent dictators}, which indicates centralisation when a blockchain is newly deployed.
\end{itemize}

\vspace{0.5em}\noindent \textbf{Known uses:}  
 \begin{itemize}
   \item \textit{Bitcoin}\footnote{\url{https://bitcoin.org/en/}\label{bitcoin}}. Satoshi Nakamoto was regarded as the \textit{benevolent dictator} of the Bitcoin ecosystem before his/her retirement.
   
   \item \textit{Ethereum}\textsuperscript{\ref{ethereum}}. The co-founder of Ethereum, Vitalik Buterin, is still active in governance-related issues of the Ethereum ecosystem.
   
   \item \textit{Tezos}\textsuperscript{\ref{tezos}}. Tezos foundation has veto power for the first twelve months after Tezos's deployment. 
 \end{itemize}

\vspace{0.5em}\noindent \textbf{Related patterns:} 

\begin{itemize}
    \item \textit{Liquid democracy.} \textit{Benevolent dictatorship} can be viewed as a variant of \textit{liquid democracy} that other stakeholders delegate their decision rights to the core developers.
\end{itemize}

\subsection{Cross-Chain Token Voting}

\vspace{0.5em}\noindent \textbf{Summary:} Specific token holders in a blockchain platform can vote for governance-related issues of another blockchain platform.

\vspace{0.5em}\noindent \textbf{Context:} Newly launched permissionless blockchain platforms may have few stakeholders to participate in the governance decision-making process.

\vspace{0.5em}\noindent \textbf{Problem:} How can a project team improve the participation rate, and ensure the security of governance decision-makings in permissionless blockchain platforms with a small community?

\vspace{0.5em}\noindent \textbf{Forces:} 

\begin{itemize}
  \item \textit{Participation rate improvement.} The participation rate should be improved. There might be few users in a newly deployed blockchain to vote for improvement proposals.

  \item \textit{Security.} Newly launched blockchain platforms should be protected from Sybil attack.
  
\end{itemize}

\begin{figure}[!ht]
	\centering
	\includegraphics[width=0.5\columnwidth]{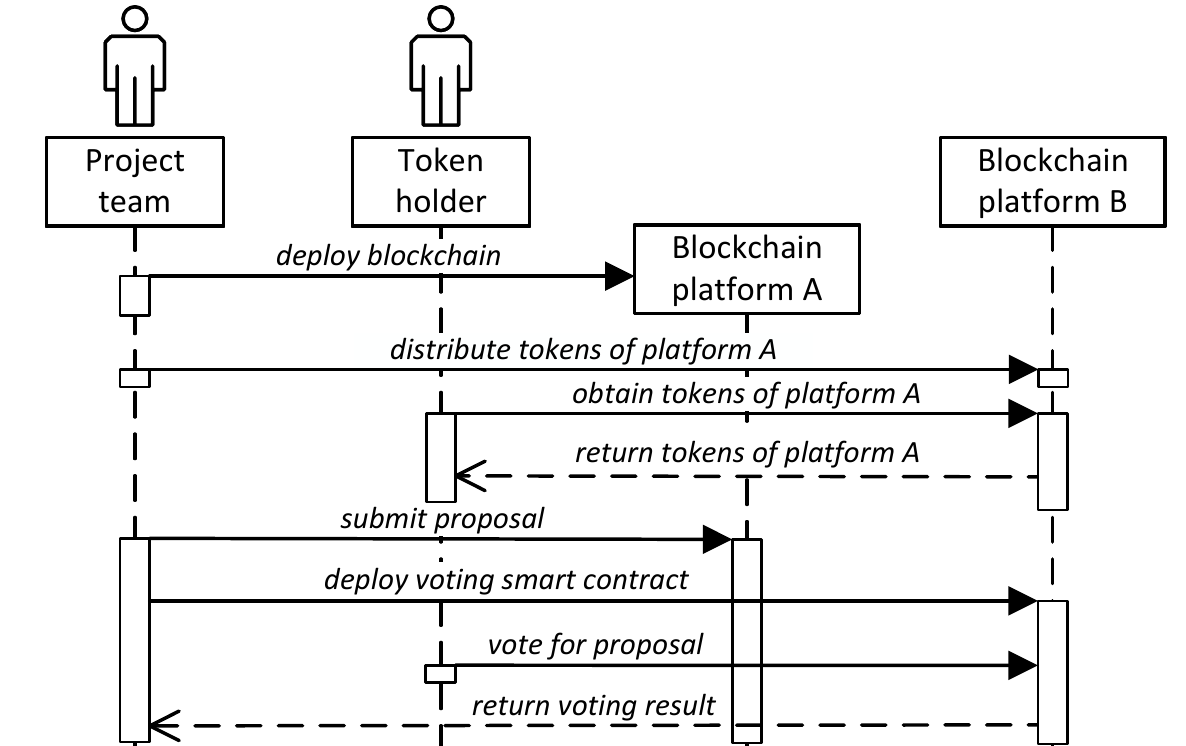}
	\caption{Cross-chain token voting.}
	\label{pic:cross_chain_voting}
\end{figure}

\vspace{0.5em}\noindent \textbf{Solution:} Fig.~\ref{pic:cross_chain_voting} depicts how to finalise a governance decision for a new blockchain platform, with the aid of other established blockchain platforms. After blockchain platform A is officially in use, its project team then issues tokens (in the name of platform A) on other mature permissionless blockchain platforms (platform B in this case). Users of platform B may invest the tokens issued by platform A. In this case, these users are considered as the token holders of platform A, even though they may not directly use platform A. These users of platform B are consequently eligible to participate in platform A's decision-making process. Specifically, when the project team submits an improvement proposal in platform A, it should develop and deploy smart contracts in platform B, to cast a vote for relevant token holders. The holders of platform A's tokens can then submit their votes for the proposal via the deployed smart contract in platform B. The result is returned to platform A's project team when the vote ends. This pattern can be considered as developers of the original blockchain platform build a decentralised application on other blockchain platforms for voting.

\vspace{0.5em}\noindent \textbf{Consequences:} 

Benefits:
\begin{itemize}
  \item \textit{Participation rate.} Stakeholders from other blockchain platforms can possess tokens of the new blockchain. Subsequently, they are eligible to participate in the decision-making process.

  \item \textit{Security.} The actual voting process happens in the selected permissionless blockchain platforms. Hence, the security is anchored to these platforms. For instance, token holders still can use their platform's accounts to vote. Their platforms may apply specific security measures.

  \item \textit{Interoperability.} Improvement proposals of the original blockchain platforms are finalised via the interoperations between multiple platforms.
\end{itemize}

Drawbacks: 
\begin{itemize}
   \item \textit{Security.} Attacks on target blockchain platforms may affect the voting results of the original blockchain.
   
   \item \textit{Cost.} There are mainly two types of cost when applying this pattern. (i) The project team needs to distribute tokens, and deploys smart contracts in other blockchain platforms, which may cost real money. (ii) The project team needs to offer incentives, attracting individuals to become token holders.
\end{itemize}

\vspace{0.5em}\noindent \textbf{Known uses:}  
 \begin{itemize}
   \item \textit{MULTAV}~\cite{MULTAV}. MULTAV is a framework deployed in IoTeX blockchain platform. IoTeX tokens are issued in Ethereum. Consequently, the IoTeX token holders in Ethereum can participate in the election of block producers in IoTeX.
   
   \item \textit{AAVE}\footnote{\url{https://aave.com/}}. AAVE supports cross-chain governance between Ethereum and Polygon.
   
   \item \textit{StakerDAO}\footnote{\url{https://www.stakerdao.com/}}. StakerDAO provides a protocol for cross-chain governance decisions. It connects multiple blockchain platforms such as Ethereum, Tezos, and Polkadot.
 \end{itemize}

\vspace{0.5em}\noindent \textbf{Related patterns:} 

\begin{itemize}

    \item \textit{Carbonvote.} \textit{Carbonvote} can be integrated with \textit{cross-chain token voting} to on-chain decision-makings.
    
    \item \textit{Quadratic voting.} \textit{Quadratic voting} can be integrated with \textit{cross-chain token voting} to on-chain decision-makings.

\end{itemize}

\subsection{Protocol Forking}

\vspace{0.5em}\noindent \textbf{Summary:} The software upgrades of a blockchain platform are implemented as forks of the blockchain.

\vspace{0.5em}\noindent \textbf{Context:} After a blockchain platform is deployed, developers upgrade the blockchain to meet new requirements.

\vspace{0.5em}\noindent \textbf{Problem:} The code of a blockchain platform is stand-alone. Any changes may affect the historical ledger contents. How can developers implement upgrades to a blockchain platform without affecting the historical ledger contents?

\vspace{0.5em}\noindent \textbf{Forces:} 

\begin{itemize}
  \item \textit{Upgradability guarantee.} Blockchain needs to be upgraded to fix software bugs, and implement new functionalities.

  \item \textit{Adaptability improvement.} Blockchain needs to adapt to the varying requirements of diverse application scenarios.
  
  \item \textit{Blockchain property preservation.} The fundamental properties of a blockchain, e.g., immutability, must not be affected by software upgrades.
\end{itemize}

\begin{figure}[!ht]
	\centering
	\subfigure[Soft fork.]{
	\centering
    \includegraphics[width=0.35\columnwidth]{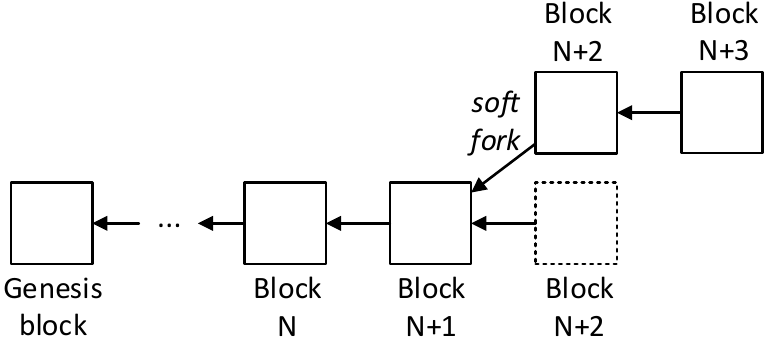}
    }
	\subfigure[Hard fork.]{
	\centering
    \includegraphics[width=0.35\columnwidth]{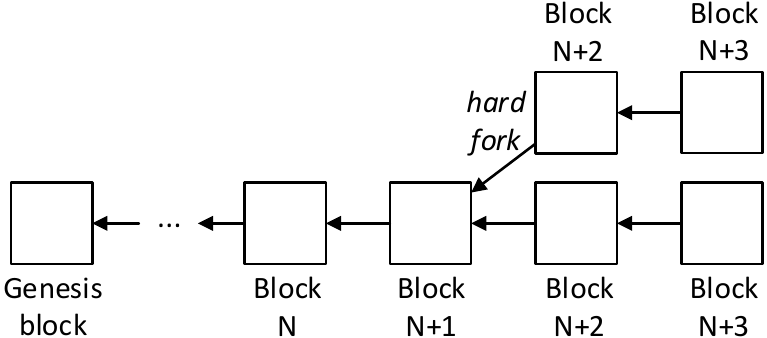}
    }
    \caption{Protocol forking.}
	\label{pic:forking}
\end{figure}

\vspace{0.5em}\noindent \textbf{Solution:} Fig.~\ref{pic:forking} is a graphical representation of \textit{protocol forking}, which can be applied to upgrade a deployed blockchain. The upgrades and evolution require a series of formalised development procedures. Upgrades accepted by the community are implemented and released as new versions via \textit{protocol forking}. Specifically, there exist two types of upgrades in blockchain: soft fork and hard fork. Soft forks mean backward-compatible upgrades, which require the majority of node operators to install the latest version of a blockchain platform. After soft fork, appending blocks need to adhere to both new and old on-chain protocols. Blocks that violate the new protocol will be abandoned by the platform (as the dotted block N+2 in Fig.~\ref{pic:forking} (a)). On the contrary, hard forks refer to the backward-incompatible upgrades, which lead to permanent divergences of a blockchain. After hard fork, a blockchain is split into two separate instances. One instance employs the old protocol while the other one follows new rules. The two instances will operate as two distinct blockchain platforms, as illustrated in Fig.~\ref{pic:forking} (b). Please note that \textit{protocol forking} in the context of blockchain governance refers to intentional software upgrades. Accidental forkings where multiple block validators intend to append new blocks at the same time are out of the scope of this paper.

\vspace{0.5em}\noindent \textbf{Consequences:} 

Benefits:
\begin{itemize}
  \item \textit{Adaptability.} \textit{Protocol forking} can improve the adaptability of a blockchain, regarding the varying requirements in diverse applications.
  
  \item \textit{Upgradability.} The project team can upgrade a blockchain platform via forking to implement new functionalities.
  
  \item \textit{Blockchain property preservation.} Forking is implemented at a certain block height. All preceding blocks of the forking block height are retained in the upgrade. Therefore, the fundamental properties of this blockchain are preserved.
\end{itemize}

Drawbacks:
\begin{itemize}
  \item \textit{Data integrity.} Hard fork may roll back certain blockchain ledger contents. All blocks succeeding the forking block height need to be discarded. In this case, data integrity is compromised. For example, a blockchain has 99 valid blocks. The project team decides to conduct a hard fork on block 80. Block 81 to 99 will then be discarded, and the next block of the forked blockchain will be new block 81.

  \item \textit{Volatility.} Hard fork may split the overall blockchain ecosystem. Stakeholders need to choose which instance they will continue to use.
  
  \item \textit{Cost.} (i) Rigorous analysis of stakeholders is needed to conduct forking; (ii) Hard fork is conducted at a specific block height, which means that the generation, broadcast, and validation of all subsequent blocks may be in vain.
\end{itemize}

\vspace{0.5em}\noindent \textbf{Known uses:}  
 \begin{itemize}
   \item \textit{Bitcoin}\textsuperscript{\ref{bitcoin}}. The Bitcoin community could not reach agreements on intended upgrades. Hence, the platform was forked into two distinct blockchain platforms: Bitcoin and Bitcoin Cash.
   
   \item \textit{Ethereum}\textsuperscript{\ref{ethereum}}. Ethereum conducted a hard fork to reverse the impacted transactions of "DAO" attack. However, part of the community refused to implement the fork, and insisted to stay on the previous version, which is subsequently known as "Ethereum Classic".
   
   \item \textit{Steem}\footnote{\url{https://steem.com/}\label{steem}}. Hive is a hard fork of Steem blockchain. It was implemented when TRON announced the takeover of Steem.
 \end{itemize}

\vspace{0.5em}\noindent \textbf{Related patterns:} 

\begin{itemize}
    \item \textit{Protocol forking} is dependent on \textit{carbonvote}, \textit{quadratic voting}, \textit{liquid democracy}, \textit{benevolent dictatorship}, and \textit{cross-chain token voting} for the acceptance of improvement proposals.
\end{itemize}

\subsection{Social Contract}

\vspace{0.5em}\noindent \textbf{Summary:} A social contract is deployed to select the future maintainers of a blockchain platform.

\vspace{0.5em}\noindent \textbf{Context:} The project team maintains the daily operation of a blockchain platform,  and modifications to the platform.

\vspace{0.5em}\noindent \textbf{Problem:} The project team may lose funding, and stop providing services to the blockchain platform. How the blockchain is maintained afterwards?

\vspace{0.5em}\noindent \textbf{Forces:} 

\begin{itemize}
  \item \textit{Maintainability guarantee.} A blockchain platform must continually have regular maintenance for normal operation.

  \item \textit{Security.} The subsequent maintainer(s) of a blockchain should be approved by the original project team.
  
  \item \textit{Transparency guarantee.} In permissionless blockchains, the transfer of maintenance rights and responsibilities should be transparent to all stakeholders, to receive their recognition.

\end{itemize}

\vspace{0.5em}\noindent \textbf{Solution:} 
The blockchain project team can deploy a \textit{social contract}, in which the requirements of how an entity becomes the eligible candidate for maintaining the blockchain platform are specified. The contract can be deployed either on-chain or off-chain, as long as it is open to all stakeholders of the platform. When the project team fails to continually provide services, the \textit{social contract} comes into effect to select the candidate maintainer(s). However, if no eligible candidate is found, the blockchain platform may have to terminate.

\vspace{0.5em}\noindent \textbf{Consequences:} 

Benefits:
\begin{itemize}
  \item \textit{Maintainability.}  When the original project team quits, the blockchain platform can be maintained by the selected candidate.
  
  \item \textit{Security.} The selection rules of candidate maintainer(s) are defined by the original project team. Therefore, the eligible maintainers are regarded as approved by the project team.

  \item \textit{Transparency.} The \textit{social contract} is open to the whole community. Therefore, the selection of candidates is monitored by all stakeholders.
\end{itemize}

Drawbacks:
\begin{itemize}
  \item \textit{Centralisation.} The selected candidate is granted decision rights of the original project team, which may cause centralisation in the future governance process when the candidate is an individual instead of a group of people.
\end{itemize}

\vspace{0.5em}\noindent \textbf{Known uses:}  
 \begin{itemize}
   \item \textit{Ethereum}\textsuperscript{\ref{ethereum}}. The Ethereum project team elaborates a \textit{social contract} in the Ethereum whitepaper: anyone possessing a certain number of ether tokens (60,102,216 * (1.198 + 0.26 * $n$), where $n$ is the number of years after the genesis block), has the right to develop a future candidate version of Ethereum. 
   
   \item \textit{Steem}\textsuperscript{\ref{steem}}. TRON announced its takeover of Steem in 2020. Afterwards, TRON's team became the maintenance team of Steem.
   
   \item \textit{Diem}\footnote{\url{https://www.diem.com/en-us/}}. Before the official deployment, Diem was sold to Silvergate. Hence, the maintenance team of Diem changed accordingly.
 \end{itemize}

\vspace{0.5em}\noindent \textbf{Related patterns:} 

\begin{itemize}
    \item \textit{Social contract} is considered related to all other patterns in this study that the contract is leveraged to select new candidate(s) as the blockchain maintainer(s), and all other patterns may be adjusted and reapplied afterwards.
    %can be applied when the blockchain is in the termination phase. The contract selects new candidate(s) to maintain the blockchain platform. The blockchain may start a new development process. All other patterns may be adjusted and reapplied. 
\end{itemize}

\subsection{Digital Signature}

\vspace{0.5em}\noindent \textbf{Summary:} Transactions are digitally signed by users, to identify the transaction sources. The digital signature can be verified by other stakeholders.

\vspace{0.5em}\noindent \textbf{Context:} Individuals generate and send transactions to blockchain to use the services.

\vspace{0.5em}\noindent \textbf{Problem:} A blockchain platform consists of multiple users and their transactions. How can stakeholders validate the origin of a transaction? Further, a transaction is broadcast by nodes. How can stakeholders ensure that a transaction is not modified during transmission?

\vspace{0.5em}\noindent \textbf{Forces:} 

\begin{itemize}
  \item \textit{Data integrity protection.} Block validators should be aware of whether a transaction has been altered by unauthorised entities.

  \item \textit{Accountability guarantee.} Users should be identifiable and accountable as the origin of transactions. They need to take responsibility for the submitted transactions.
  
  \item \textit{Non-repudiation guarantee.} A user cannot deny that he/she has sent specific transactions to blockchain, to avoid their responsibility for the transactions.
\end{itemize}

\begin{figure}[!ht]
	\centering
	\includegraphics[width=0.38\columnwidth]{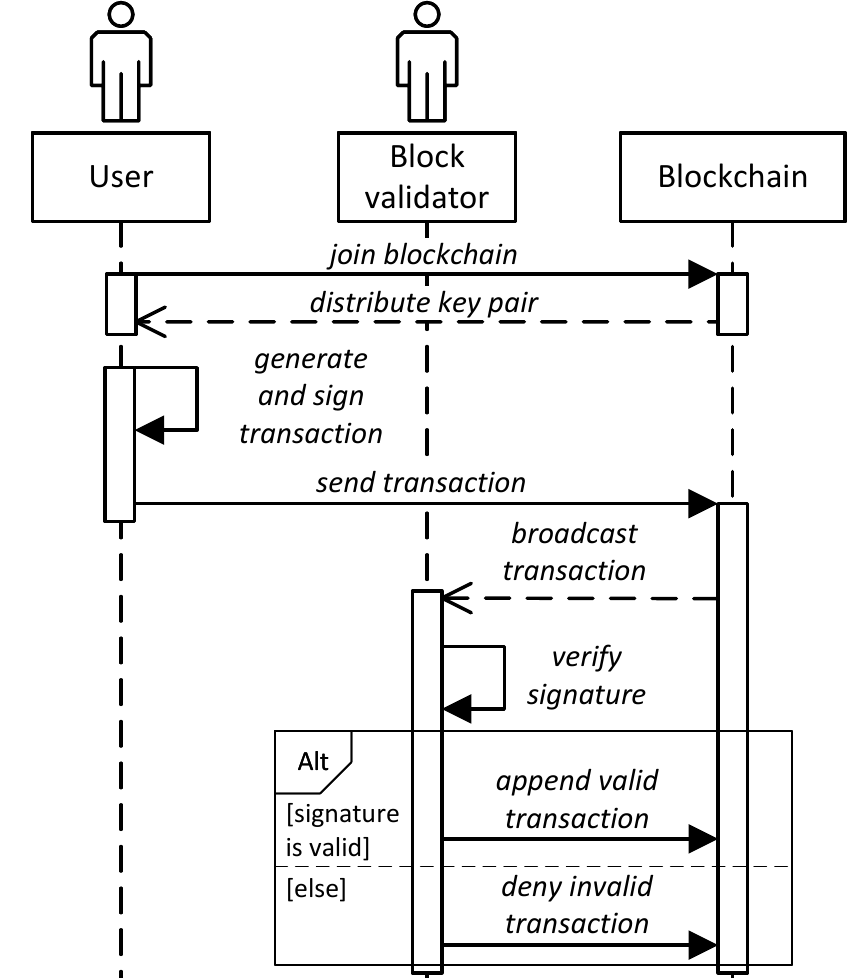}
	\caption{Digital signature.}
	\label{pic:digital_signature}
\end{figure}

\vspace{0.5em}\noindent \textbf{Solution:} A simplified \textit{digital signature} process adopted from~\cite{hashizume2009digital} is depicted in Fig.~\ref{pic:digital_signature}. When a user joins a blockchain platform, the underlying decentralised public key infrastructure (DPKI) distributes a key pair to the user. The private key is kept by the user himself/herself, while the public key is transparent to all other stakeholders. When the user generates transactions, he/she should sign the transactions before sending them to blockchain. The signature is generated by hashing the original contents, and encrypting the hash value via user's private key. The submitted transactions are broadcast across the blockchain network. Subsequently, a block validator can verify whether the transaction is valid via the signature. The validator can hash the transaction contents, and decrypt the signature using user's public key. The obtained hash value is then compared with the generated hash value, to check whether the transaction is modified. Valid transactions will be stored in a block and appended to blockchain, while invalid transactions are denied and discarded.

\vspace{0.5em}\noindent \textbf{Consequences:} 

Benefits:
\begin{itemize}
  \item \textit{Data integrity.} The integrity of transaction contents is protected by the employed hashing algorithm. Any revision to original contents will result in a different hash value, hence, verification failure.

  \item \textit{Accountability.} \textit{Digital signature} can help trace the accountable individuals in real-world in permissioned blockchains, and responsible blockchain addresses in permissionless blockchains.

  \item \textit{Non-repudiation.} A user cannot deny sending particular transactions as he/she must sign before submitting.
\end{itemize}

Drawbacks: 
\begin{itemize}
   \item \textit{Accountability.} Although transactions are digitally signed by users, it is still hard to identify accountable real-world identities for illegal behaviours, due to the inherent anonymity in permissionless blockchains. There is no mapping between users' real-world identities and on-chain public keys.
   
   \item \textit{Security.} If the private key is lost/compromised, a user then loses control of the corresponding public key. The attacker can forge blockchain transactions using the compromised private key.
\end{itemize}

\vspace{0.5em}\noindent \textbf{Known uses:} This pattern is universally utilised in existing blockchain platforms, e.g., Ethereum\textsuperscript{\ref{ethereum}}, Bitcoin\textsuperscript{\ref{bitcoin}}, and Corda\footnote{\url{https://www.corda.net/}}, etc. A user needs to set up a password, which is used to generate the key pair. Each transaction is signed by the private key, and can be verified using the corresponding blockchain address (public key).

\vspace{0.5em}\noindent \textbf{Related patterns:} 

\begin{itemize}
    \item \textit{Transaction filter.} An unsigned transaction should be rejected by the filter.
    
    \item \textit{Key management}~\cite{SSIpattern}. The key management patterns can assist users to protect private keys.
\end{itemize}

\subsection{Transaction Filter}

\vspace{0.5em}\noindent \textbf{Summary:} A filter can be utilised to examine the submitted transactions, to ensure the validity of transaction format/content.

\vspace{0.5em}\noindent \textbf{Context:} Transactions are the data entries of blockchain. Individuals use blockchain services via transactions.

\vspace{0.5em}\noindent \textbf{Problem:} How can the platform ensure that a submitted transaction can engage in the subsequent validation process?

\vspace{0.5em}\noindent \textbf{Forces:} 

\begin{itemize}
  \item \textit{Usability guarantee.} A transaction should meet the format/content requirements before being stored in the transaction pool and broadcast around network.

  \item \textit{Security.} Avoid a malicious user from sending unauthorised or harmful information to blockchain. Specifically,  in a permissioned blockchain for a specific use, irrelevant data should not be stored on-chain. For instance, in a blockchain-based supply chain application~\cite{XU2019399}, a freight yard examiner should upload only freight yard examination results. Other information is unauthorised.
\end{itemize}

\begin{figure}[!ht]
	\centering
	\includegraphics[width=0.5\columnwidth]{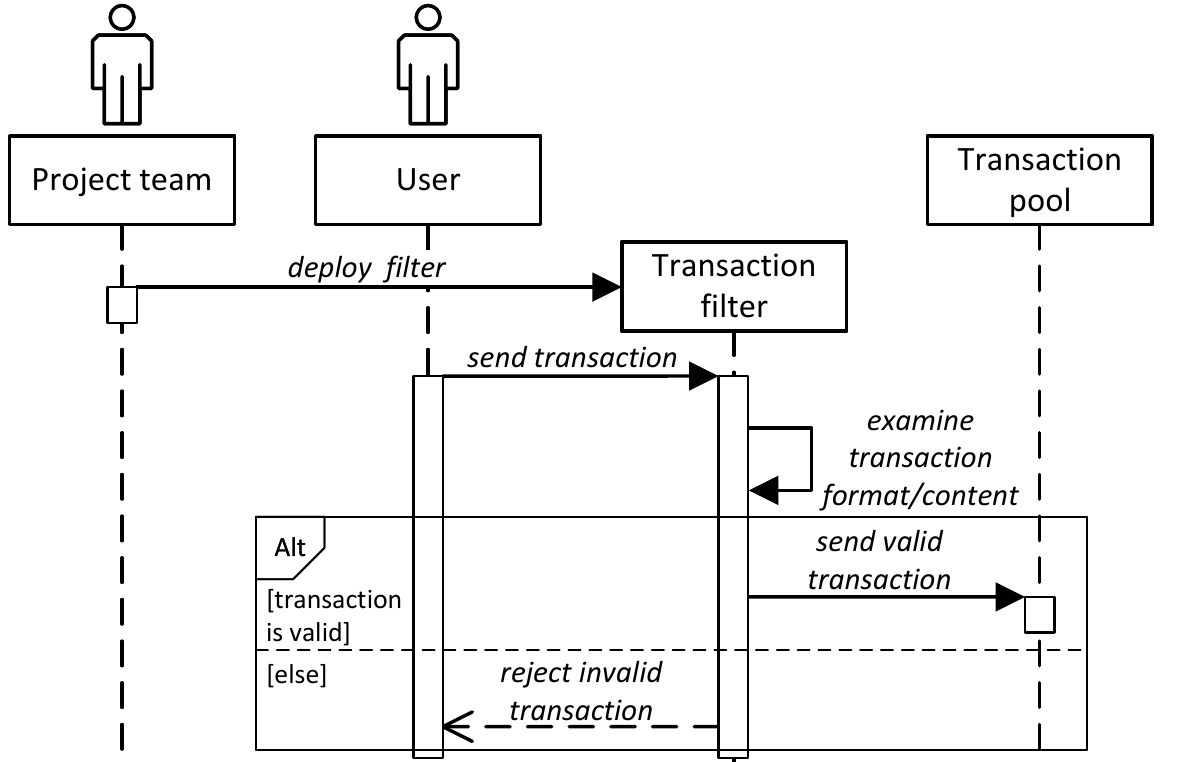}	
	\caption{Transaction filter.}
	\label{pic:transaction_filter}
\end{figure}

\vspace{0.5em}\noindent \textbf{Solution:} As shown in Fig.~\ref{pic:transaction_filter}, a \textit{transaction filter} can be deployed by the blockchain project team. When users generate and send transactions to blockchain, the filter checks the format and the carried contents of received transactions based on predefined settings. Only the valid transactions can be stored in a node's transaction pool for the subsequent process, while others are rejected and removed from blockchain.

\vspace{0.5em}\noindent \textbf{Consequences:} 

Benefits:
\begin{itemize}
  \item \textit{Usability.} Invalid transactions are rejected, which ensures all accepted transactions are usable for the subsequent broadcast and validation process.
  
  \item \textit{Security.} \textit{Transaction filter} can detect whether there is unauthorised or harmful information in a transaction. Invalid transactions are rejected from being stored on blockchain.
  
  \item \textit{Performance.} \textit{Transaction filter} removes invalid transactions from blockchain, which can facilitate the subsequent transaction validation process.
\end{itemize}

Drawbacks: 
\begin{itemize}
  \item \textit{Security.} Malicious users may send the hash values or encrypted text of illegal information to blockchain. It is hard to detect this type of harmful information. Consequently, such information may still be stored in blockchain (in the form of hash value or encrypted text), although it is not allowed.

   \item \textit{Privacy.} Individuals' privacy may be leaked when examining transactions. This may be impacted by applying other techniques such as data mining.
   
\end{itemize}

\vspace{0.5em}\noindent \textbf{Known uses:}  
 \begin{itemize}
   \item Many blockchain platforms (e.g., Ethereum\textsuperscript{\ref{ethereum}}, Bitcoin\textsuperscript{\ref{bitcoin}}, and Hyperledger Indy\footnote{\url{https://www.hyperledger.org/use/hyperledger-indy}}) have their own definition and setting of transaction format.
   
   \item IBM\footnote{\url{https://www.ibm.com/docs/en/wip-bs?topic=SSCG66/iot-blockchain/developing/event_filtering.html}}. IBM pinpointed that when filtering identical and similar transactions, only state changes are sent to blockchain to reduce blockchain traffic.
   
   \item MultiChain\footnote{\url{https://www.multichain.com/}}. MultiChain deploys Smart Filter, which enables custom transactions and data rules via checking the inputs, outputs, and metadata of transactions.
 \end{itemize}

\vspace{0.5em}\noindent \textbf{Related patterns:} 

\begin{itemize}
    \item \textit{Digital signature.} An unsigned transaction should be rejected by the filter.
    
    \item \textit{Data migration}~\cite{data_migration}. If malicious information is stored on-chain, data migration via hard fork can be conducted to reverse the relevant transactions.
\end{itemize}

\subsection{Log Extractor}

\vspace{0.5em}\noindent \textbf{Summary:} Log extractor allows application providers to extract DApp usage information from blockchain.

\vspace{0.5em}\noindent \textbf{Context:} Application providers develop DApps and provide different functionalities through smart contracts, while users interact with DApps by sending transactions.

\vspace{0.5em}\noindent \textbf{Problem:} How can application providers understand the actual usage of a DApp (i.e., when, where,
and under what circumstances the DApp is used by whom)?

\vspace{0.5em}\noindent \textbf{Forces:} 

\begin{itemize}
  \item \textit{Functional suitability guarantee.} Application providers need to ensure that the actual usage of a DApp is as the intended usage to provide complete, correct, and appropriate services.

  \item \textit{Adaptability improvement.} Application providers need to understand and analyse the usage of DApp, to perform suitable adaptations.
  
  \item \textit{Accountability guarantee.} Application providers need to know who are using the deployed DApp and related usage information, to trace accountable user for exceptions.
\end{itemize}

\begin{figure}[!ht]
	\centering
	\includegraphics[width=0.5\columnwidth]{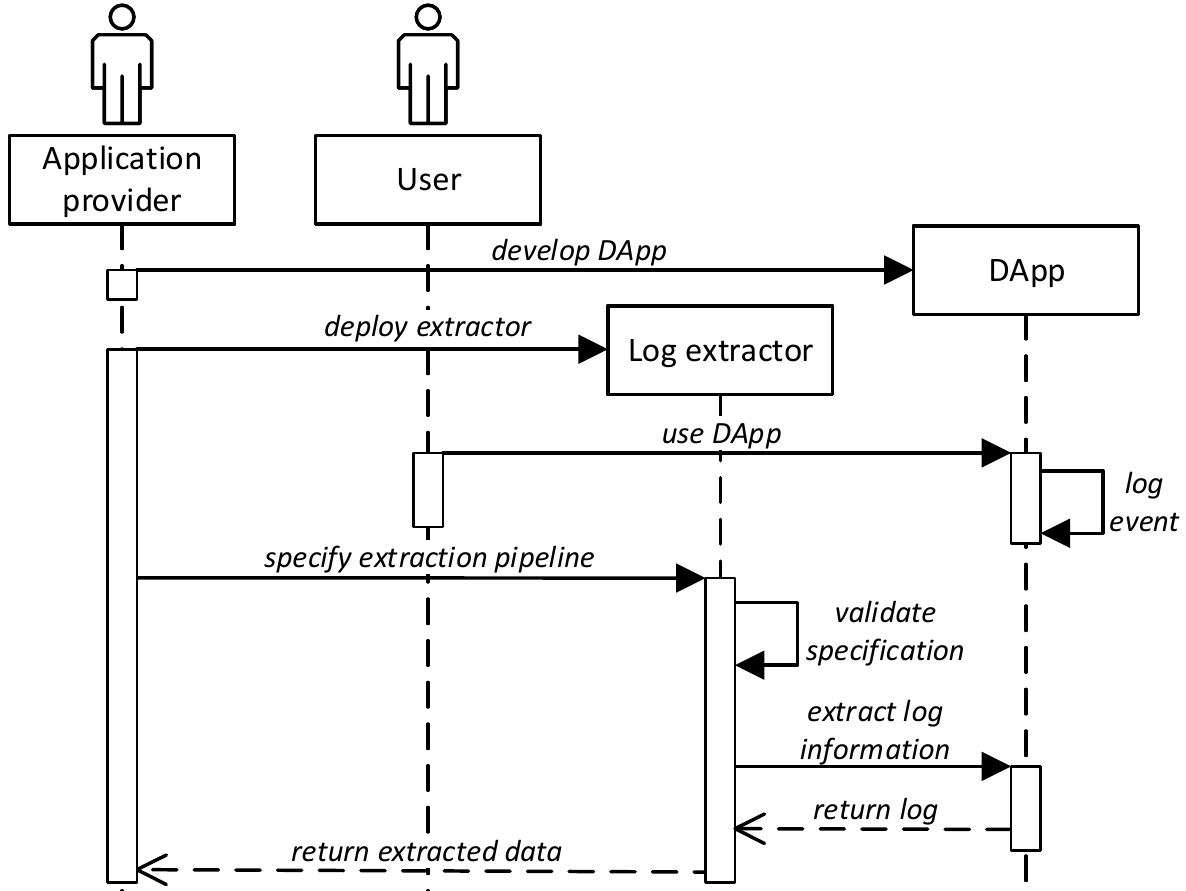}
	\caption{Log extractor.}
	\label{pic:log_extractor}
\end{figure}

\vspace{0.5em}\noindent \textbf{Solution:} Fig.~\ref{pic:log_extractor} depicts how can application providers obtain the logs of a DApp via a \textit{log extractor}. After a DApp is deployed on blockchain, individuals can use it via transactions. For specific operations, the DApp can record the events according to codified rules, which will provide usage logs to application providers. Providers can build up the pipeline in the log extractor, which is then validated to ensure a correct format of usage information. The extractor is connected to blockchain, and extracts target event logs block by block. The extracted data is then transformed, and returned to application providers for further analysis.

\vspace{0.5em}\noindent \textbf{Consequences:} 

Benefits:
\begin{itemize}
  \item \textit{Functional suitability.} Through the extracted logs, application providers can learn how the DApp is operated, and compare the actual usage with intended functionalities.

  \item \textit{Adaptability.} According to the actual usage logs, application providers can analyse and decide whether the DApp should adapt to new requirements.
  
  \item \textit{Accountability.} Extracted logs record who conduct certain operations to the DApp, which can be used to identify accountable users.
\end{itemize}

Drawbacks:
\begin{itemize}
  \item \textit{Privacy.} In permissionless blockchains, on-chain events are transparent, and can be extracted by all stakeholders. This may leak the usage information of DApps.
\end{itemize}

\vspace{0.5em}\noindent \textbf{Known uses:}  
 \begin{itemize}
   \item \textit{Blockchain Logging Framework}~\cite{blockchain_logging_framework, Mining_Blockchain_Processes}. Researchers have applied process mining techniques to extract data from blockchain. Currently, the proposed framework supports Ethereum and Hyperledger.
   
    \item \textit{BlockSLaaS}~\cite{BlockSLaaS}. BlockSLaaS can provide logs to cloud forensic investigators for forensic investigation in cloud computing.
   
   \item \textit{BlockStore}~\cite{BlockStore}. BlockStore can assign storage to renters. The storage ownership is logged in blockchain, where users can extract and verify.
 \end{itemize}

\vspace{0.5em}\noindent \textbf{Related patterns:} 

\begin{itemize}
    \item \textit{Scam list.} Suspicious blockchain addresses can be tagged and alarmed to all stakeholders after analysing extracted logs.
\end{itemize}

\section{Conclusion}
\label{conclusion}

Blockchain governance is significant to preserve software qualities, to provide a trustworthy blockchain ecosystem to the community. In this study, we present 14 architectural patterns for the design of governance-driven blockchain systems. This pattern language can provide guidance for practitioners to perform blockchain governance.

In addition to the patterns analysed in this paper, further research can explore how existing patterns are related to blockchain governance. For instance, data migration patterns for the removal of on-chain data~\cite{data_migration}, patterns for blockchain-based applications~\cite{xu2018pattern}, and different consensus mechanisms\cite{consensus_survey, PoW_pattern}. Besides, we plan to explore the architecture decisions that are related to blockchain governance.

\section*{Acknowledgments}

We thank Professor Eduardo Fernandez for the helpful comments and guidance in the shepherding process.

% Generated by IEEEtran.bst, version: 1.14 (2015/08/26)

%\bibliographystyle{IEEEtran}
%\bibliography{bibliography}

%%
%% If your work has an appendix, this is the place to put it.

\end{document}